# PROPOSAL OF A CARSHARING SYSTEM TO IMPROVE URBAN MOBILITY

## Fernando ALMEIDA
Faculty of Engineering of Oporto University, INESC TEC, Rua Dr. Roberto Frias s/n, 4200-465, Porto, Portugal
almd@fe.up.pt

## Pedro SILVA
Higher Polytechnic Institute of Gaya, Av. dos Descobrimentos 333, 4400-103, V.N.Gaia, Portugal
ispg3881@ispgaya.pt

## João LEITE
Higher Polytechnic Institute of Gaya, Av. dos Descobrimentos 333, 4400-103, V.N.Gaia, Portugal
ispg3756@ispgaya.pt

**Abstract**
Carsharing is a model of renting vehicles for short periods of time, where the payment is made according to the time and distance effectively traveled. Carsharing offers a simple, economical and smart alternative to urban mobility, that is already being adopted in the major cities in the world. The proposed methodology consisted in the development of a decision support system that simplifies the process of choosing carsharing services. Adopting the AHP method, the user can indicate their preferences in the choice of vehicles, and the system returns an ordered list of the most suitable available vehicles based on their geographic location. The findings of the project indicate that the use of this system encourage and simplify the use of carsharing services, which will allow to enhance the financial, mobility and environment advantages inherent to their use.

## 1. INTRODUCTION

The main objective of carsharing is to give people who do not have their own vehicle the possibility of using a car without the costs of their acquisition, obligations and legal responsibilities. Carsharing is the concept of renting a car similar to the "car renting" system, but for a shorter period of time. Danielis et al. (2017) state that cars can be rented by individuals or organizations (e.g., commercial business, a public agency or a cooperative).

During last decades the urban transport and mobility systems of our cities have changed significantly. At first, the increasing use of private transport in industrialized countries provided greater accessibility. However, and in the long term, it may result in serious negative issues, such as pollution, congestion problems and excessive consumption of energy (Jorge & Correia, 2013). Therefore, we assist today to a high interest by car producers and private investors in carsharing services, which emerged by the increasing awareness of public stakeholders to the environmental impact of urban policies, the large amount of funds available in the past decades for innovative solutions in urban mobility, and the maturity level reached by ICT technologies and solutions.

There are generally three models of carsharing services. The first carsharing model is the traditional (two-way) or station based, based on the idea that shared vehicles must be picked and returned to the same station. A slight evolution of this model is the one way carsharing model, which allows users to initiate and end the rental period in different stations. The last model that appeared in the market is the free floating one. In this case, members can pick up and return the vehicle in any parking lot within the operation area, paying only for the driving time (Ferrero et al., 2015).

This research proposes the development of a support decision system that can improve the process of choosing carsharing services based on the user preferences. The paper is organized as follows: we initially perform a revision of literature in the field of decision support systems. After that initial phase, we present the adopted methodology by detailing the used methods, functional and non-functional requirements, and the physical and logical architecture of the application. Then, we present the prototype, followed by an analysis of the results, attending to the potential financial and environmental benefits, and the scalability of the application. Finally, the conclusions of this work are drawn.

## 2. LITERATURE REVIEW

### 2.1. Concept and structure of a decision support system

Decision Support Systems (DSS) are computer-based information systems, especially designed to help managers in decision making problems. They are conceivable to automate large decision making processes, using computer-based software that analyzes huge amounts of data. It helps companies, institutions and governments reduce costs, increase profitability and enhance the quality (Tripathi, 2011).

According to Asemi et al. (2011) and Gupta & Singhal (2013) a DSS has a number of common characteristics, namely: (i) provides support for decision maker mainly in semi structured and unstructured situations; (ii) attempts to improve the effectiveness of decision-making; (iii) can handle large amount of data; (iv) focuses in a very domain specific; (v) operates as an interactive system; and (vi) supports optimization and heuristic approach.

The basic structured of a DSS is formed of three related components: the database, the model design and the interface. A database consists of a set of internal or external consistent data organized in a physical location, such as a file or a computer (Elmasri and Navathe, 2015). The model design is a set of mathematical representations through variables and constraints are able to represent a real life situation. These models can be used in descriptive, optimized or heuristic problems (Sharda et al., 2014). The user interface is a component that provides the communication between the user and the decision support system. Marakas (2002) and Stanciu (2009) highlights the importance of the user interface, by referring that it should be simples, consistent and flexible, because it is the component responsible to interact with the end user.

### 2.2. Decision support systems in carsharing services

Decision support systems have been used and implemented successfully in a wide range of different areas, such as industry (Bakhrankova, 2010), lumber (Acosta & Corral, 2017), government (Peignot et al. 2013), health (Assena et al, 2013), social sciences (Hillegersberg & Koenen, 2016) or education (Fakeeh, 2015). The application of a DSS for carsharing services is not so widely spread, but we may already find some relevant studies in this specific field.

Agapitou et al. (2014) looks to the implementation of car-sharing in different continents and proposes a number of ways to improve the effectiveness of carsharing. This study also demonstrates a significant number of successful implementations of different carsharing models. Other authors, such as Litman (2000) and Schuster et al. (2005) confirm that carsharing has been observed to have a positive impact on urban mobility. Mitchell et al. (2010) explain that shared vehicles can have much higher utilization rates that single-user private vehicles because each vehicle spends more time on the road rather than parked.

Cepolina and Farina (2012) provide an overview of relocation strategies of carsharing services to avoid the need of advance reservations of vehicles. Another relevant overview in the same field is provided by Agatz et al. (2012) that systematically outline the optimization challenges that arise when developing technology to support ride-sharing and survey the related operations research models in the academic literature. In the same direction, Nourinejad (2014) provides an overview about location issues of carsharing services and presents two dynamic models for real-time management of carsharing and ridesharing services.

The majority of studies in carsharing services proposes mainly different strategies for the implementation of carsharing services. Nourinejad and Roorda (2014) propose a dynamic optimization-simulation model as a decision support system for one-way carsharing organizations. Cepolina and Farina (2011) present a new shared vehicle system for urban areas based on a fleet of eco-sustainable Personal Intelligent City Accessible Vehicles (PICAVs). Sonneberg et al. (2015) propose a decision support system for the optimization of electric carsharing stations. The idea is to present an optimization model that anticipates optimal locations of stations, which will conduct to a higher profitability.

## 3. APPROACH & METHODOLOGY

### 3.1. Functional and non-functional requirements

According to Sommerville (2015) a functional requirement is defined as "a condition or a capability with which the system must agree". In this sense, functional requirements specify actions that a system must be able to perform without regard to physical constraints. Therefore, they specify the input and output behavior of a system.

The following groups of functional requirements were considered in the context of our project.

- User authentication - the registration of the user in the application is done by completing a registration form. After the email has been confirmed by the application, the user can enter his/her access credentials;

- Search for carsharing services - the car sharing services will be listed according to the user's geographic area;

- Simulate total costs and contract a service - a simulation of the cost of the service chosen by the user will be made. Several factors are considered to estimate the total price, such as the cost per minute (travelling), cost per minute (in standby), travel time duration, parking time interval, and travel distance range. Finally, the service can be contracted by the user;

- Choice of service based on user preferences - the AHP method is used to capture the user preferences for the carsharing service. The user can specify the importance given to various factors, such as vehicle consumption, comfort and safety;

- Perform the service evaluation - the user can evaluate his or her satisfaction with the rented vehicle. The evaluation is made adopting a scale from 1 to 5, related to comfort, consumption and safety. Each evaluation is recorded individually for each vehicle. The average of the total evaluation is calculated, and it is presented as an indicator of the overall performance of each vehicle.

Sommerville (2015) also defines non-functional requirements as "constraints on the services or functions offered by the system such as timing constraints, constraints on the development process, standards, etc. Unlike functional requirements, these non-functional requirements are not explicitly exposed by the client, but must be implicitly understood by the developer.

In the context of our project, it is pertinent to consider the following non-functional requirements:

- Usability - the system offers a very clean and intuitive interface, preventing the insertion in the system of invalid and redundant content by the user;

- Efficiency - the efficiency of the application has been taken into account, particularly looking to response time the application takes to calculate the best service, according to user preferences;

- Security - to ensure that, in case of any failure occurrence, whether natural or human, customer data is stored in backup systems. Furthermore, user passwords are stored encrypted.

### 3.2. Methods

One of the basis of a DSS system is the inclusion of models. The idea consists in the adoption of interactive computer-based systems that help decision makers utilize data and models to solve unstructured problems. The AHP method, proposed by Saaty (1980), helps structure the decision-makers' thoughts and can help in organizing and structuring the problem in a manner that is simple to follow and analyze. The main goal of AHP is helping in structuring the complexity, measurement and synthesis of rankings. To make a good decision, the decision maker must know and define: the problem, the need and purpose of the decision, the criteria and sub criteria to evaluate the alternatives, the stakeholders and groups affected (Russo & Camanho, 2015).

The process of operation of AHP method can be explained as follows: (i) decomposition of the problem into a hierarchy of goal, criteria, sub-criteria and alternative; (ii) collection of data from experts or decision-makers corresponding to the hierarchic structure; (iii) pairwise comparisons of the various criteria generated at step 2 and its sub sequential organization into a square matrix; (iv) comparison of the principal eigenvalue and the corresponding normalized right eigenvector of the comparison matrix; (v) evaluation of the matrix consistency; (vi) multiplication of the rating of each alternative by the weights of the sub-criteria to get local ratings with respect to each criterion. Then, the local ratings are multiplied by the weights of the criteria and aggregated to get global ratings (Bhushan and Rai, 2004).

### 3.3. Physical and logical architecture

The physical architecture presents the technologies used in the system and describes the way that the various technological elements are structured in layers.

The physical architecture is divided into three layers. The database (layer 1) contains the saved data from the application; the business logic (layer 2) is based on C# programming language and it is responsible to interconnect the information stored in the database with what is presented to the user; finally, the presentation layer (layer 3) is implemented using the Windows Presentation Foundation (WPF) technology. The physical architecture is illustrated in the figure 1.

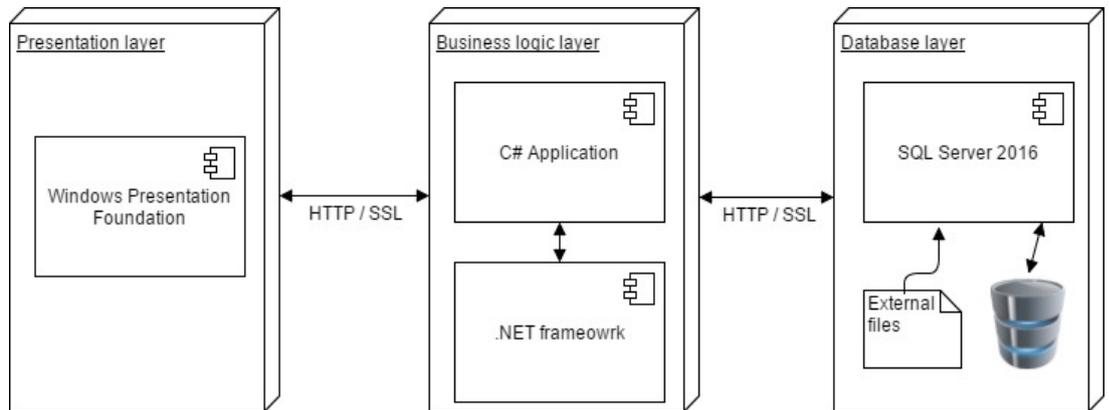

FIGURE 1 - PHYSICAL ARCHITECTURE OF THE APPLICATION

On the other side, the logical architecture tells us how the various classes of the application are related. It is also here that the model of the database shall be presented and described.

The logical architecture of the system is divided into seven modules. The initial module of the system is the "login" component. After validating the user's credentials, the user access to the "listing" component where he/she can visualize all the available vehicles. The "listing" component is associated with four other components: "simulation", "contract", "internal evaluation" and "external evaluation". The "simulation" module uses the "AHP method" class that is responsible for the implementation the AHP method in C#. The "external evaluation" module uses two external files. The "import_aval.xml" contains external evaluations and "impor_aval.xsd" specifies the structure of the xml file. The logical architecture is depicted in figure 2.

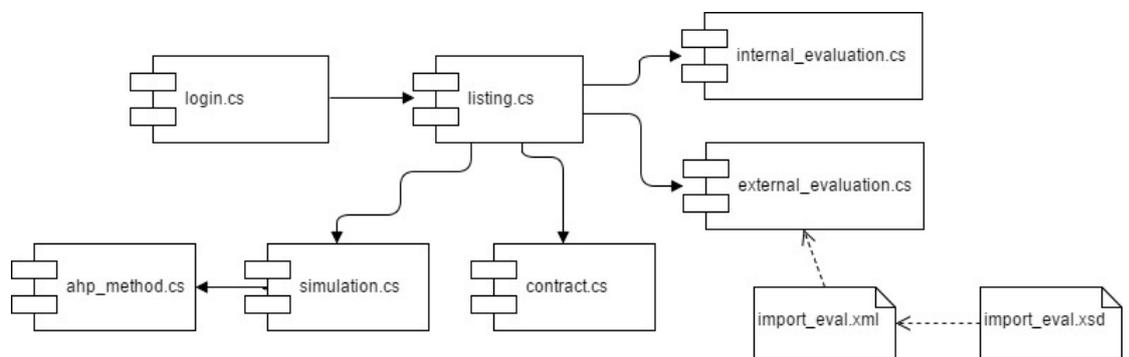

FIGURE 2 - LOGICAL ARCHITECTURE OF THE APPLICATION

Finally, the database is normalized into the third normal form (3NF). The class diagram has a total of 15 tables. The main tables of our architecture are responsible to store data regarding clients, partners, vehicles and services. Only one table in the database is not interconnected with others, because it contains information regarding authentication.

## 4. PROTOTYPE

The initial interface of the application is the login window. The user must specify the access credentials or press the "create account" button if he/she does not have an account created. This scenario is illustrated in figure 3.

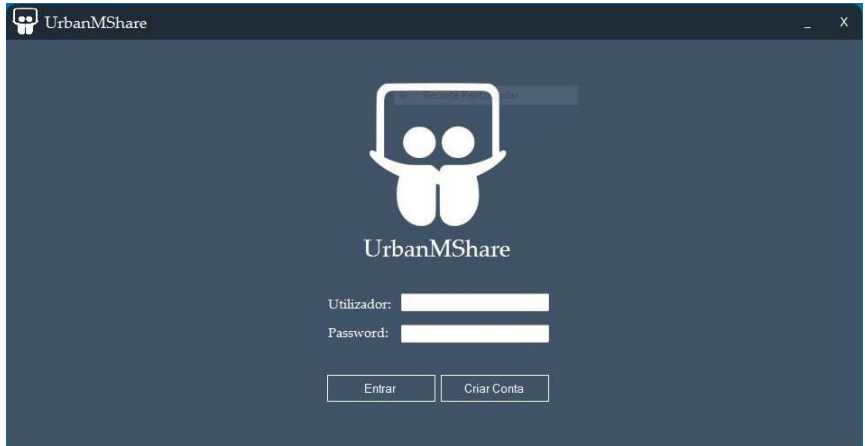

FIGURE 3 - INITIAL INTERFACE OF THE APPLICATION

After the login process has been successfully executed, the user access to the main panel of the frontend. In the main menu the user visualizes the list of cars in the user's geographical. For each available car we show information regarding the brand, model, color, existence of air conditioning, price per hour, type of fuel, number of kilometers (kms), and location where the vehicle is parked. The average of the evaluations according to each parameter is also provided. This situation is illustrated in figure 4.

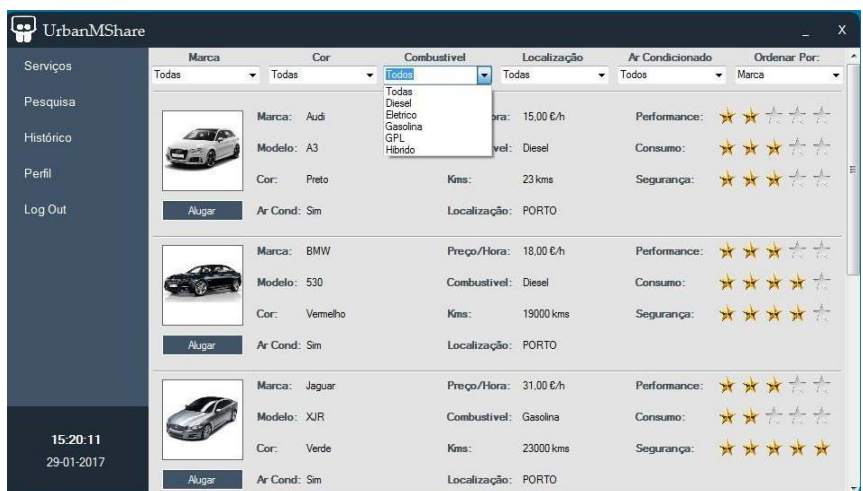

FIGURE 4 - LISTING OF AVAILABLE SERVICES

Finally, it is relevant to show the interface of the choice of service based on user preferences. Using the interface shown in figure 5, the user can specify the importance given to criteria such as performance, consumption and security.

FIGURE 5 - CHOICE OF SERVICES BASED ON USER PREFERENCES

## 5. RESULTS & DISCUSSIONS

The analysis of the potentialities offered by the carsharing application is carried out taking into account simultaneously external and internal perspectives. At the external level, we estimate the financial benefits and environment of their use. At the internal level, we analyze the scalability of the system, considering both the increase in the number of vehicles and criteria considered when performing the AHP method.

### 5.1. Estimation of financial and environmental benefits

The financial impact by the introduction of carsharing services is estimated by changing the average number of trips per month (ANM) and the average duration of each trip in hours (ADT). Additionally, we calculate the amount of savings in dollars for month (SM) and year (SY). For that, we use the simulator provided by Zipcar[1] to perform these calculations. Table I provides an overview about the financial impact of the adoption of carsharing. The following profile of costs per month with own vehicles were considered: car payment ($300); insurance ($30); gas ($100); license, registration and taxes ($30); maintenance ($30); and parking ($60). The results confirm that the financial impact of carsharing services is higher when we have lower ANM. Within the same ANM the financial benefits decrease with the increment of ADT. Finally, it is important to refer that for scenario IV we have two situations with negative financial performance. It is the case when the ANM is equal to 12 and ADT has the value of 7 and 8, which represents a situation where exists a significant level of carsharing journeys per month with long durations.

---

[1] http://www.zipcar.com/is-it

TABLE I - ESTIMATION OF FINANCIAL IMPACT BY THE INTRODUCTION OF CARSHARING

| Scenario I | | | | | | | |
|---|---|---|---|---|---|---|---|
| **ANM** | 1 | 1 | 1 | 1 | 1 | 1 | 1 | 1 |
| **ADT** | 1 | 2 | 3 | 4 | 5 | 6 | 7 | 8 |
| **SM ($)** | 542 | 534 | 527 | 519 | 511 | 503 | 501 | 494 |
| **SY ($)** | 6504 | 6408 | 6324 | 6228 | 6132 | 6036 | 6012 | 5928 |
| **Scenario II** | | | | | | | |
| **ANM** | 4 | 4 | 4 | 4 | 4 | 4 | 4 | 4 |
| **ADT** | 1 | 2 | 3 | 4 | 5 | 6 | 7 | 8 |
| **SM ($)** | 519 | 494 | 466 | 438 | 410 | 383 | 355 | 327 |
| **SY ($)** | 6228 | 5928 | 5592 | 5256 | 4920 | 4596 | 4260 | 3924 |
| **Scenario III** | | | | | | | |
| **ANM** | 8 | 8 | 8 | 8 | 8 | 8 | 8 | 8 |
| **ADT** | 1 | 2 | 3 | 4 | 5 | 6 | 7 | 8 |
| **SM ($)** | 494 | 438 | 383 | 327 | 286 | 234 | 181 | 128 |
| **SY ($)** | 5928 | 5256 | 4596 | 3924 | 3432 | 2808 | 2172 | 1536 |
| **Scenario IV** | | | | | | | |
| **ANM** | 12 | 12 | 12 | 12 | 12 | 12 | 12 | 12 |
| **ADT** | 1 | 2 | 3 | 4 | 5 | 6 | 7 | 8 |
| **SM ($)** | 466 | 383 | 313 | 234 | 155 | 76 | -3 | -82 |
| **SY ($)** | 5592 | 4596 | 3756 | 2808 | 1860 | 912 | -36 | -984 |

The environmental impact can be estimated looking to the Greenhouse Gases (GHG). This indicator includes the following gases: carbon dioxide ($CO_2$), methane ($CH_4$), nitrous oxide ($N_2O$) and fluorinated gases.

According to Nijland et al. (2015), which conducted a study to analyze the impact of car sharing on mobility and $CO_2$ emissions in the Dutch territory, the following conclusions can be extracted:

▪ Car shares drive around 15% to 20% fewer car kilometers than before they started carsharing;

▪ Car shares emit between 175 and 265 fewer kilograms of $CO_2$ per person, per year, due to their reduced car ownership. The reduction of $CO_2$ is around 8% to 13% compared to car ownership and car use.

Another study conducted by Martin and Shaheen (2016) in the US and Canada territories confirms a significant reduction in terms of GHG. For that, they look to the GHG emissions in five cities (i.e., Calgary, San Diego, Seattle, Vancouver and Washington). The range of vehicles removed by the introduction of carsharing services was bigger for the cities of Seattle and Calgary. On the other side, the estimated % reduction in GHG was more significant for the cities of Vancouver (-15%) and Washington (-18%).

### 5.2. Scalability of the system

A relevant discussion point in a project based on information technologies is to test and measure its scalability. According to Bondi (2014), scalability is the system's ability to handle a high amount of processes or the potential to increase processing in order to deal with the growth of tasks. Basically, it refers to the ability of a system to increase its total output when there is an increase in load. The scalability must be thought at the design time of the application, because it is not a simple feature that can be added later in the application. Therefore, the decisions that we make during coding and in early stages, strongly affects the scalability potential of an application.

In our application, we establish two types of scalability:

▪ Horizontal scalability - it includes the introduction of several comparison criteria in the AHP method;

▪ Vertical scalability - it includes the addition of new vehicles from the carsharing companies.

The horizontal scalability of the system is conditioned by the number of comparison criteria. Presently we have only used three criteria: (i) performance; (ii) consumption; and (iii) security). Apparently, according to a theoretical model, the system would be perfectly horizontally scalable, since adding a new criterion to the model would only mean building a new matrix for that criterion and adding a new line to the matrix of comparison of the importance of alternatives. However, in practical terms, this situation deserves more attention, since the addition of a new criterion also implies the comparison of this new criterion to the others, and this process must be carried out by the user. Since usability is one of the most important criteria in the development of an information system, then it must be ensured that the user is able to effectively and intuitively use the system (Hoo & Jaafar, 2013). Furthermore, and according to Saaty (2003), to serve both consistency and redundancy to the AHP method, it is best to keep the number of criteria at seven.

In order to estimate the vertical scalability of the system we proceeded to an empirical study that consisted in increasing the number of alternatives, as it is depicted in figure 6.

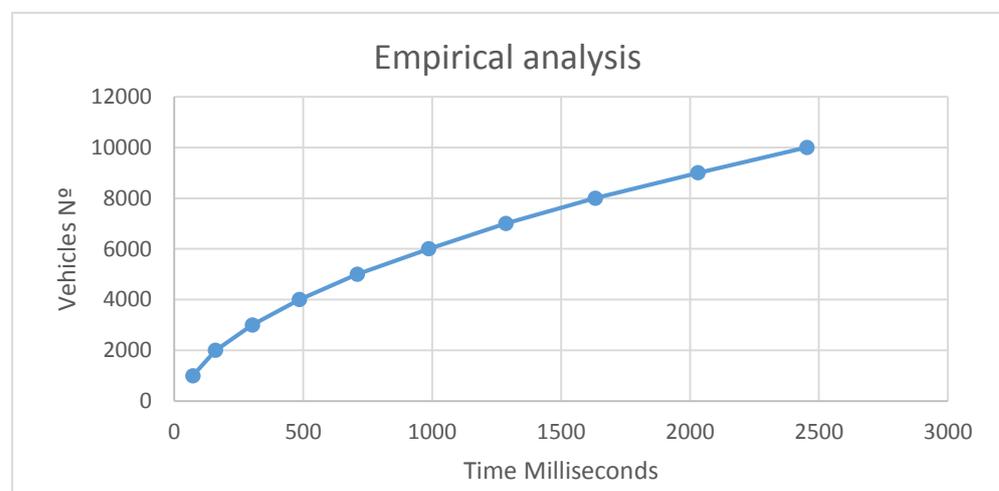

FIGURE 6 - ESTIMATION OF VERTICAL SCALABILITY

The execution time raises approximately exponentially as the number of the vehicle increases. The growth rate is much lower than when we have a larger number of vehicles. For example, an addition of thousand vehicles to the model (from 1000 to 2000) implies an increase of the execution time in approximately 90 milliseconds. However, the same increase (from 9000 to 10000) causes an increase in the execution time by approximately 400 milliseconds.

## 6. CONCLUSIONS

Carsharing services are becoming a viable alternative to private vehicles due to their multiple benefits. The carsharing system offers the advantage of having a car without the responsibilities, costs and maintenance that a particular car requires. Econometric analysis reveals that carsharing services are essentially advantageous when the number of trips per month and their duration is low.

For cities, carsharing brings two large groups of advantages: (i) impact on mobility; and (ii) impact on the environment. In terms of mobility, the decrease in the number of cars in circulation contributes to an improvement in city traffic, as well as greater efficiency in the use of public space. On the other hand, in environmental terms, the decrease of the number of vehicles in the city causes a reduction of the pollution with impact on the improvement of the quality of air, reduction of greenhouse gases and reduction of the energetic dependence.

The application developed under this project offers to citizens a support mechanism in the process of choosing carsharing services, which will potentially increase the use of carsharing services and, consequently, will contribute to improving urban mobility. The application developed suggests to the user the best available carsharing services according to his/her specific needs and it also allows the simulation of the total costs of a service before its reservation.

As future work we consider relevant the evolution of the application to a mobile environment. Furthermore, it would also be interesting if the feedback regarding the quality of the service of each use was also incorporated in the choice of carsharing services, in order to benefit the services with better scores.